\documentclass[aps,prb,groupedaddress,amsmath,eqsecnum,floatfix,twocolumn]{revtex4}
\usepackage{graphicx,amssymb}
\usepackage{bm}
\usepackage{amsmath}
\def\beq{\begin{equation}}
\def\eeq{\end{equation}}
\def\beqa{\begin{eqnarray}}
\def\eeqa{\end{eqnarray}}
\def\zero{{(0)}}
\def\one{{(1)}}
\def\two{{(2)}}

\def\N{\Gamma}
\newcommand{\nn}{\nonumber\\}

\newcommand{\mo}{\langle\sigma\rangle}
\newcommand{\mt}{\langle\sigma^2\rangle}
\newcommand{\mth}{\langle\sigma^3\rangle}

\begin{document}
\title{Depletion potential in the infinite dilution limit}
\author{Santos Bravo Yuste}
\email{santos@unex.es} \homepage{http://www.unex.es/fisteor/santos/}
\author{Andr\'es Santos}
\email{andres@unex.es} \homepage{http://www.unex.es/fisteor/andres/}
\affiliation{Departamento de F\'{\i}sica, Universidad de
Extremadura, E-06071 Badajoz, Spain}
\author{Mariano L\'opez de Haro}
\email{malopez@servidor.unam.mx}
\homepage{http://xml.cie.unam.mx/xml/tc/ft/mlh/} \affiliation{Centro
de Investigaci\'on en Energ\'{\i}a, Universidad Nacional Aut\'onoma
de M\'exico (U.N.A.M.), Temixco, Morelos 62580, M{e}xico}
\date{\today}
\begin{abstract}
The depletion force and depletion potential between two in principle
unequal ``big'' hard spheres embedded in a multicomponent mixture of
``small'' hard spheres are computed using the Rational Function
Approximation method for the structural properties of hard-sphere
mixtures [S. B. Yuste, A. Santos, and M. L\'opez de Haro, J. Chem.
Phys. {\bf 108}, 3683 (1998)]. The cases of equal solute particles
and of one big particle and a hard planar wall in a background
 monodisperse hard-sphere fluid are explicitly analyzed. An
improvement over the performance of the Percus--Yevick theory and
good agreement with available simulation results are found.

\end{abstract}

\maketitle

\section{Introduction}
\label{sect_1}

Excluded volume interactions  in hard-sphere mixtures are
interesting for a number of reasons. For one thing, colloidal
systems are often modeled as mixtures of dissimilar hard spheres
and, experimentally, it is known that this interaction plays an
important role in the observed behavior.  Moreover, due only to the
existence of a disparate size ratio between solute and solvent, the
(purely entropic) effect on the solvent mediated interaction forces
between solute particles in a hard-sphere suspension may be quite
dramatic. Take for instance the case of two (not necessarily equal)
big hard spheres immersed in a fluid of small hard spheres. When the
distance between the two big spheres is less than the diameter of
the small spheres, the latter may not got into the gap. This
depletion effect induces an imbalance in the local pressure leading
ultimately to an effective attraction between the big spheres. A
rather similar phenomenon  occurs when, in the presence of say a
hard planar wall, one has a single big hard sphere (solute) in a sea
of small hard spheres (solvent).

{The concept of depletion force was} first introduced by Asakura and
Oosawa\cite{AO54}  over fifty years ago in the context of
colloid-polymer mixtures. Ever since, a great number of papers
devoted to depletion interactions involving hard-sphere mixtures
have appeared in the literature. The approaches have also been
varied ranging from experiment,\cite{YLCDVK01} density functional
theory,\cite{GED98,GRDDE99,BRLRD99,RED00,AAM03,ZLM03,LM04,RK05}
computer simulations,\cite{BBF96,DAS97,DvRE99,AL01,LXM01,HH06,MK07}
virial expansions,\cite{MCL95} the Derjaguin
approximation,\cite{HJ02,OL04,O04,HH07} or the integral equation
formulation of liquid state
theory.\cite{AAM03,HL86,H88,AP90,H92,MK00,HWT01,AL02,HWT03,CRM03,GMA04,AMA05,CRM06}

Very recently we have addressed the problem of deriving the density
profiles of multicomponent hard-sphere mixtures near a planar hard
wall\cite{MYSL07} using an alternative approach to the integral
equation theory for the structural properties of the
system.\cite{YSH98,HYS07} The main aim of this paper is to
complement our former results with the study of the depletion
potential for various systems involving hard-sphere mixtures.
Specifically we will deal with the depletion interaction between two
different (big) hard spheres immersed in a multicomponent mixture of
(small) hard spheres. A limiting case of this situation is when the
diameter of one of the big spheres becomes infinite and so this
sphere is seen as a hard planar wall both by the other big sphere
and by the multicomponent mixture of small spheres. It should be
emphasized that our approach represents an improvement over the
Percus--Yevick (PY) theory. {In fact, it retains the major asset of
the latter, namely it also yields analytical results in Laplace
space. Further, our approach is technically simple too, so that the
improvement over the PY theory is not achieved at the expense of
adding difficulty to the theoretical development.}

The paper is organized as follows. In the next section we provide a
relatively simple derivation of the Asakura--Oosawa depletion
potential by looking at the  exact results for  the radial
distribution functions of a multicomponent hard-sphere mixture to
first order in density. This is followed in Section \ref{sec3} by a
summary of the results for the structural properties of a
multicomponent hard-sphere mixture obtained by using the Rational
Function Approximation (RFA) method.\cite{YSH98,HYS07} This section
also includes the limit where two of the species in the mixture (the
solute) are present in vanishing concentration and the special cases
where the solute particles have equal diameters or one is seen as a
wall both by the other solute particle and by the solvent. Section
\ref{sec4} presents the results  both of the depletion potential and
the depletion forces for some illustrative cases and compares them
with available computer simulation data.\cite{DAS97,AL01,LM04,MK07}
We close the paper in Section \ref{sec5} with a discussion of the
results and some concluding remarks.

\section{Radial distribution functions to first order in density: the Asakura--Oosawa result}
\label{sec2}
In this section we present a simple derivation of the
Asakura--Oosawa depletion potential\cite{AO54} that follows from
exact results of the structural properties of multicomponent
hard-sphere mixtures rather than the original geometrical arguments.
Let us consider an $(N+2)$-component hard-sphere mixture in which
the closest distance between a sphere of species $i$ and a sphere of
species $j$ is $\sigma_{ij}$. Thus, $\sigma_i\equiv \sigma_{ii}$ can
be considered as the ``diameter'' of a sphere of species $i$. The
mixture can be either additive, i.e.,
$\sigma_{ij}=(\sigma_{i}+\sigma_{j})/2$, or nonadditive, i.e.,
$\sigma_{ij}\neq(\sigma_{i}+\sigma_{j})/2$. The number density of
the mixture and the mole fraction of species $i$ will be denoted by
$\rho$ and $x_i$, respectively. To first order in $\rho$, the radial
distribution functions $g_{ij}(r)$ are exactly given by
\beq
g_{ij}(r)=g_{ij}^{(0)}(r)+\rho g_{ij}^{(1)}(r)+\mathcal{O}(\rho^2),
\label{8AO}
\eeq
where
\beq
g_{ij}^{(0)}(r)= \Theta(r-\sigma_{ij}),
\label{9AO.0}
\eeq
\beq
g_{ij}^{(1)}(r)=\sum_{k=1}^{N+2} x_k g_{ij;k}^{(1)}(r),
\label{9AO}
\eeq
\beqa
g_{ij;k}^{(1)}(r)&=&\frac{\pi}{12r}[r^2+2(\sigma_{ik}+\sigma_{kj})r-3(\sigma_{ik}-\sigma_{kj})^2]
\nn &&\times
(\sigma_{ik}+\sigma_{kj}-r)^2\Theta(r-\sigma_{ij})\Theta(\sigma_{ik}+\sigma_{kj}-r).
\label{11AO}
\eeqa
In Eqs.\ \eqref{9AO.0} and \eqref{11AO} $\Theta(x)$ is the Heaviside
step function. A derivation of Eq.\ \eqref{11AO} can be found in
Appendix \ref{appA}.

Now we assume that the mole fractions of species $i=N+1$ and $i=N+2$
(here labeled as $i=a$ and $i=b$, respectively) vanish, so that the
other species ($i=1,2,\ldots,N$) constitute the solvent. In that
case, the depletion potential $u_{ab}(r)$ for the effective
interaction between the solute spheres $a$ and $b$ is given by
\beq
\beta
u_{ab}(r)=-\ln g_{ab}(r),
\label{1AO}
\eeq
where $\beta\equiv 1/k_B T$, $k_B$ being the Boltzmann constant and
$T$ the absolute temperature.  According to Eqs.\ \eqref{8AO} and
\eqref{9AO.0}, to first order in $\rho$ (and for $r>\sigma_{ab}$),
one  has
\beq
\beta
u_{ab}(r)=-\rho g_{ab}^{(1)}(r),
\label{2AO}
\eeq
so that, taking Eqs.\ \eqref{9AO} and \eqref{11AO} into account,
\beqa
\beta u_{ab}(r)&=&-\rho\frac{\pi}{12r}\sum_{i=1}^N
x_i(\sigma_{ai}+\sigma_{bi}-r)^2[r^2+2(\sigma_{ai}\nn
&&+\sigma_{bi})r-3(\sigma_{ai}-\sigma_{bi})^2]\Theta(\sigma_{ai}+\sigma_{bi}-r).\nn
\label{3AO}
\eeqa

If both solute spheres are identical ($\sigma_{ai}=\sigma_{bi}$,
$\sigma_{ab}=\sigma_a$), Eq.\ \eqref{3AO} becomes
\beqa
\beta u_{aa}(d)&=&-\rho\frac{\pi}{12}\sum_{i=1}^N
x_i(2\sigma_{ai}-\sigma_{a}-d)^2(d+\sigma_a+4\sigma_{ai})\nn
&&\times\Theta(2\sigma_{ai}-\sigma_a-d),
\label{4AO}
\eeqa
where we have defined the distance $d=r-\sigma_{a}$. If,
furthermore, the $ai$ interaction is additive, namely
$2\sigma_{ai}=\sigma_a+\sigma_i$, then
\beq
\beta u_{aa}(d)=-\rho\frac{\pi}{12}\sum_{i=1}^N
x_i(\sigma_{i}-d)^2(d+3\sigma_a+2\sigma_{i})\Theta(\sigma_{i}-d).
\label{5AO}
\eeq
This result coincides with the Asakura--Oosawa
expression.\cite{AO54,DAS97}

We now go back to the case $a\neq b$, define $z=r-\sigma_{ab}$ and
assume that the $ab$ and $bi$ interactions are additive. In the
limit $\sigma_b\to\infty$ the sphere $b$ becomes a wall and Eq.\
\eqref{3AO} reduces to
\beqa
\beta u_{wa}(z)&=&-\rho\frac{\pi}{6}\sum_{i=1}^N
x_i\left(\sigma_{ai}+\frac{\sigma_i-\sigma_a}{2}-z\right)^2\nn
&&\times(2z+\sigma_a+4\sigma_{ai}-\sigma_i)\nn
&&\times\Theta\left(\sigma_{ai}+\frac{\sigma_i-\sigma_a}{2}-z\right).
\label{6AO}
\eeqa
Again, if furthermore the $ai$ interaction is also additive,
\beq
\beta u_{wa}(z)=-\rho\frac{\pi}{6}\sum_{i=1}^N
x_i\left(\sigma_i-z\right)^2(2z+3\sigma_a+\sigma_i)\Theta\left(\sigma_i-z\right).
\label{7AO}
\eeq
This result also  coincides with the corresponding Asakura--Oosawa
expression.\cite{AO54,DAS97}

Note that the validity of Eq.\ \eqref{3AO} to first order in $\rho$
actually extends to \emph{any} interaction among the solvent
particles, {including  the so-called Asakura--Oosawa model
($\sigma_i=0$, $\sigma_{ai}>\sigma_a/2$, $\sigma_{bi}>\sigma_b/2$)},
i.e., only the solute-solvent ($ai$ and $bi$) and solute-solute
($ab$) interactions need to be those of hard spheres.
 We must also point out
that Eq.\ \eqref{3AO}, while applying to first order in density
only, is quite general in the following sense: (i) the solvent may
be in general polydisperse, (ii) the solute-solvent and
solute-solute hard-sphere interactions are not necessarily additive,
and (iii) the two solute spheres may have arbitrary sizes. We remark
that, in general, the depletion potential $u_{ab}(r)$ is not a
polynomial function of distance but a polynomial of degree four
divided by the distance between the centers of spheres $a$ and $b$.
Only in the cases $a=b$ [see Eq.\ \eqref{4AO}] and $b\to\text{wall}$
[see Eq.\ \eqref{6AO}] does the potential become a polynomial (of
degree three).

 The results of this section are exact but restricted to a low-density solvent.
{In particular, the Asakura--Oosawa potentials turn out to be purely
attractive (with  a range corresponding to the diameter of the
solvent  particles) and scale with the solvent density. Neither of
these features remains as the density is increased. While it would
be nice to have some measure of the error made in using Eqs.\
\eqref{5AO} or \eqref{7AO} in actual situations, there is
unfortunately no clear-cut way to estimate such an error. Instead,
we note that, according to the qualitative discussion  performed in
Ref.\ \onlinecite{DAS97}, the entropic force grows faster than the
bulk density and becomes repulsive for distances on the order of
half  the diameter of the solvent particles. Hence, in order to
account for these and other finite-density effects on the depletion
interaction one must adopt a different strategy and resort to
approximations. In the next section we present our analytical
approach, which includes the PY approximation as a particular case.}

\section{The Rational Function Approximation Method}
\label{sec3} In this section we {start by recalling} the main
aspects of the RFA method for multicomponent hard-sphere
mixtures\cite{YSH98} and refer the interested reader to our recent
review paper\cite{HYS07} and references therein for details.

As in the preceding section, let us consider an $(N+2)$-component
fluid of  hard spheres of diameters $\sigma_i$ and mole fractions
$x_i$ ($i=1,\ldots,N+2$). Now we restrict ourselves to the additive
case, but otherwise the density $\rho$ is arbitrary. The packing
fraction of the mixture is $\eta=(\pi/6)\rho\langle
\sigma^3\rangle$, where
\beq
\langle \sigma^n\rangle=\sum_{i=1}^{N+2} x_i \sigma_i^n
\label{n1}
\eeq
denotes the $n$th moment of the size
distribution. According to the RFA,\cite{YSH98,HYS07} the Laplace
transform $G_{ij}(s)$ of $rg_{ij}(r)$ is given by
\beq
\label{3.6}
G_{ij}(s)=\frac{e^{-\sigma_{ij} s}}{2\pi s^2} \left[{\sf L}(s)\cdot
{\sf B}^{-1}(s)\right]_{ij},
\eeq
where
$\sigma_{ij}=(\sigma_i+\sigma_j)/2$ and ${\sf L}(s)$ and ${\sf
B}(s)$ are $(N+2)\times (N+2)$ matrices given by
\beq
\label{3.7}
L_{ij}(s)=L_{ij}^\zero+L_{ij}^\one s+L_{ij}^\two s^2,
\eeq
\beq
{B}_{ij}(s)= (1+\alpha s)\delta_{ij}-{A}_{ij}(s),
\label{1}
\eeq
\beqa
\label{3.8}
A_{ij}(s)&=&\rho x_i\left[\varphi_2(\sigma_i s)\sigma_i^3
L_{ij}^\zero +\varphi_1(\sigma_i s)\sigma_{i}^2
L_{ij}^\one\right.\nn&&\left. +\varphi_0(\sigma_{i} s)\sigma_{i}
L_{ij}^\two\right].
\eeqa
In Eq.\ (\ref{3.8}),
\beq
\label{2.9}
\varphi_n(x)\equiv x^{-(n+1)}\left(\sum_{m=0}^n \frac{(-x)^m}{m!}-
e^{-x}\right).
\eeq

By construction, Eq.\ (\ref{3.6}) complies with the requirement
$\lim_{s\rightarrow\infty}s
e^{\sigma_{ij}s}G_{ij}(s)=\text{finite}$. Further,  the coefficients
of $s^0$ and $s$ in the power series expansion of $s^2 G_{ij}(s)$
must be 1 and 0, respectively. This  allows us to express ${\sf
L}^\zero$ and ${\sf L}^\one$ in terms of ${\sf L}^\two$ and
$\alpha$:
\beq
\label{3.13}
L_{ij}^\zero=\lambda+\lambda'\sigma_j+2\lambda'\alpha-
\lambda\rho\sum_{k=1}^{N+2} x_k\sigma_k L_{kj}^\two,
\eeq
\beq
\label{3.14}
L_{ij}^\one=\lambda\sigma_{ij}+\frac{1}{2}\lambda'\sigma_i\sigma_j
+(\lambda+\lambda'\sigma_i)\alpha-\frac{1}{2}\lambda\rho\sigma_i
\sum_{k=1}^{N+2} x_k\sigma_k L_{kj}^\two,
\eeq
where
\beq
\lambda\equiv \frac{2\pi}{1-\eta},\quad
\lambda'\equiv \frac{6\pi\eta}{(1-\eta)^2}\frac{\langle \sigma^2\rangle}{\langle \sigma^3\rangle}.
\label{2}
\eeq

In principle, ${\sf L}^\two$ and $\alpha$ can be chosen arbitrarily
without violating any basic physical condition. In particular, the
choice $L_{ij}^\two=\alpha=0$ gives the PY solution.\cite{L64,BH77}
Since we want to go beyond this approximation, we will determine the
coefficients ${\sf L}^\two$ and $\alpha$ by taking prescribed values
for $g_{ij}(\sigma_{ij} )$ and the associated thermodynamically
consistent  (reduced) isothermal compressibility $\chi$. {Hence, in
our case,}
\beq
\label{3.17}
{L_{ij}^\two}={2\pi\alpha\sigma_{ij}}g_{ij}(\sigma_{ij} )
\eeq
and  $\alpha$ is found to be the smallest real root of an algebraic
equation.

Here we will take for $g_{ij}(\sigma_{ij})$ the accurate extended
Carnahan--Starling--Kolafa (eCSK3) approximation\cite{MYSL07,SYH05}
\beqa
g_{ij}(\sigma_{ij})&=&\frac{1}{1-\eta}+\frac{3 \eta}{2
\left(1-\eta\right)^2}\frac{\mt}{\mth}\frac{\sigma_i
\sigma_j}{\sigma_{ij}}+\frac{\eta^2(5-2\eta+2\eta^2)}{12(1-\eta)^3}\nn
&&\times \left(\frac{\mt}{\mth}\frac{\sigma_i
\sigma_j}{\sigma_{ij}}\right)^2+\frac{\eta^2(1+\eta)}{6(1-\eta)^2}\left(\frac{\mt}{\mth}\frac{\sigma_i
\sigma_j}{\sigma_{ij}}\right)^3,\nn
\label{eCSK3}
\eeqa
which is thermodynamically consistent with the (reduced) isothermal
compressibility  $\chi$ derived from Boubl\'{\i}k's equation of
state,\cite{Bou86} namely
\beqa
1/\chi&=&\frac{1}{(1-\eta)^2}+\frac{6\eta}{(1-\eta)^3}\frac{\mo\mt}{\mth}.\nn
&&+\eta^2\frac{27-8\eta-8\eta^2+4\eta^3}{3(1-\eta)^4}
\frac{\mt^3}{\mth^2}.
\label{n2}
\eeqa
In the case where one of the species (say $j=N+2$) becomes a wall
(i.e., $x_{N+2}\to 0$, $\sigma_{N+2}\to\infty$), Eq.\ \eqref{eCSK3}
reduces to
\beqa
g_{wi}(\sigma_{wi})&=&\frac{1}{1-\eta}+\frac{3 \eta}{
\left(1-\eta\right)^2}\frac{\mt}{\mth}{\sigma_i
}+\frac{\eta^2(5-2\eta+2\eta^2)}{3(1-\eta)^3}\nn
&&\times\left(\frac{\mt}{\mth}{\sigma_i
}\right)^2+\frac{4\eta^2(1+\eta)}{3(1-\eta)^2}\left(\frac{\mt}{\mth}{\sigma_i
}\right)^3,\nn
\label{eCSK3w}
\eeqa

\subsection{Infinite dilution of species $a$ and $b$}
Now we assume that the mole fractions of  species $i=N+1$ and
$i=N+2$ (labeled again as $i=a$ and $i=b$, respectively) vanish,
i.e., $x_a\to 0$, $x_b\to 0$. In that case, those species do not
contribute to the total packing fraction or to other average values:
\beq
\langle \sigma^n\rangle\to \sum_{i=1}^N x_i \sigma_i^n, \quad n\leq 3.
\label{3}
\eeq
We assume that this is the case, even if the diameters of the
spheres of species $a$ and $b$ are infinitely larger than those of
the solvent species.

The limits $x_a\to 0$, $x_b\to 0$ imply that the last two rows of
the $(N+2)\times (N+2)$ matrix  ${\sf A}$ defined by Eq.\
\eqref{3.8} vanish, so that the $(N+2)\times (N+2)$ matrix
$\mathsf{B}$ defined by Eq.\ \eqref{1} has the following block
structure:
\beq
\mathsf{B}= \left(
\begin{array}{ccc|cc}
B_{11}&\cdots &B_{1N}& -A_{1a}&-A_{1b}\\
B_{21}&\cdots &B_{2N}& -A_{2a}&-A_{2b}\\
\vdots&\ddots &\vdots&\vdots&\vdots\\
B_{N1}&\cdots &B_{NN}&-A_{Na}&-A_{Nb}\\
\hline
0&\cdots& 0&1+\alpha s&0\\
0&\cdots& 0&0&1+\alpha s
\end{array}
\right).
\label{5}
\eeq
Analogously,
\beq
\mathsf{B}^{-1}= \left(
\begin{array}{ccc|cc}
(\mathsf{B}^{-1})_{11}&\cdots &(\mathsf{B}^{-1})_{1N}&
C_{1a}&C_{1b}\\
(\mathsf{B}^{-1})_{21}&\cdots
&(\mathsf{B}^{-1})_{2N}&C_{2a}&C_{2b}\\
\vdots&\ddots &\vdots&\vdots&\vdots\\
(\mathsf{B}^{-1})_{N1}&\cdots
&(\mathsf{B}^{-1})_{NN}&C_{Na}&C_{Nb}\\
\hline
0&\cdots& 0&\frac{1}{(1+\alpha s)}&0\\
0&\cdots& 0&0&\frac{1}{(1+\alpha s)}
\end{array} \right),
\label{6}
\eeq
where
\beq
C_{ia}\equiv \frac{1}{1+\alpha s}\sum_{k=1}^N \left({\sf
B}^{-1}\right)_{ik}A_{ka},\quad i=1,\ldots,N,
\label{10}
\eeq
with a similar expression for $C_{ib}$.

Insertion of Eq.\ \eqref{6} into Eq.\ \eqref{3.6} gives $G_{aa}(s)$,
$G_{ab}(s)$, $G_{bb}(s)$, $G_{ai}(s)$ and $G_{bi}(s)$ for
$i=1,\ldots,N$, and $G_{ij}(s)$ for $i,j=1,\ldots, N$. The latter
quantities refer to  the $N$-component mixture solvent and, as
expected, are not affected by the presence of the solute particles
$a$ and $b$. The solute-solvent correlation functions $G_{ai}(s)$
are
\beq
 G_{ai}(s)=\frac{e^{-\sigma_{ai}s}}{2\pi s^2} \sum_{j=1}^N
L_{aj}\left({\sf B}^{-1}\right)_{ji},\quad i=1,\ldots,N.
\label{14}
\eeq
These quantities have been considered elsewhere\cite{MYSL07} in the
wall limit $\sigma_a\to\infty$. Here we want to focus on the
solute-solute correlations $G_{ab}(s)$ in the presence of the
$N$-component bath. The result can be written as
\beqa
 G_{ab}(s)&=&\frac{1}{1+\alpha s}\left[\frac{e^{-\sigma_{ab}s}}{2\pi s^2}{L_{ab}(s)}\right.\nn
 &&\left.+ \sum_{i=1}^N
e^{\frac{1}{2}(\sigma_i-\sigma_b)s}G_{ai}(s)A_{ib}(s)\right],
\label{15}
\eeqa
where Eqs.\ (\ref{10}) and (\ref{14}) have been used. Equation
\eqref{15} is the main result of this section and readily allows us
to get both  the depletion potential $u_{ab}(r)$ and the depletion
force $F_{ab}(r)$. They are given by
\beqa
\beta u_{ab}(r)&=&-\ln g_{ab}(r)\nn &=&-\ln
\frac{\mathcal{L}^{-1}[G_{ab}(s)]}{r},
\label{phi}
\eeqa
\beqa
\beta F_{ab}(r)&=& -
\beta\frac{du_{ab}(r)}{dr}=\frac{g'_{ab}(r)}{g_{ab}(r)}\nn
&=&\frac{\mathcal{L}^{-1}[s
G_{ab}(s)-\sigma_{ab}e^{-\sigma_{ab}s}g_{ab}(\sigma_{ab}^+)]}{\mathcal{L}^{-1}[G_{ab}(s)]}-\frac{1}{r}
,\nn
\label{Fdep}
\eeqa
where $\mathcal{L}^{-1}$ denotes the inverse Laplace transform
operator. In Eqs.\ \eqref{phi} and \eqref{Fdep} it is understood
that $r>\sigma_{ab}$ since both the potential and the force are of
course singular in the region $0\leq r\leq \sigma_{ab}$. We recall
that the PY results are recovered by setting $\alpha=0$.

When the $N$-component mixture solvent becomes a pure fluid
 (i.e.\ $\sigma_i=\sigma_1$ for $i=1,\ldots,N$), one has
$L_{ij}=L_{11}$ and $A_{ij}=x_i A_1$ for $i,j= 1,\ldots,N$, and
$A_{ia}=x_i A_a$ and $A_{ib}=x_i A_b$ for $i=1,\ldots,N$, where
\beqa
\label{3.8.1}
A_m(s)&=&\rho\left[\varphi_2(\sigma_1 s)\sigma_1^3 L_{1m}^\zero
+\varphi_1(\sigma_1 s)\sigma_{1}^2 L_{1m}^\one\right.\nn &&\left.
+\varphi_0(\sigma_{1} s)\sigma_{1} L_{1m}^\two\right],\quad m=1,a,b.
\eeqa
 In that case, Eqs.\ (\ref{14}) and (\ref{15}) become
\beq
G_{a1}(s)=\frac{e^{-\sigma_{a1} s}}{2\pi
s^2}\frac{L_{a1}({s})}{1+\alpha s - A_1(s)},
\label{18}
\eeq
\beq
G_{ab}(s)=\frac{e^{-\sigma_{ab} s}}{2\pi
s^2}\frac{L_{ab}(s)\left[1+\alpha
s-A_1(s)\right]+L_{a1}({s})A_{b}(s)}{(1+\alpha s)\left[1+\alpha s -
A_1(s)\right]}.
\label{19}
\eeq

 In what follows we will
consider the particular cases in which the sizes of the two solute
spheres are the same or when one of the solute spheres has an
infinite size so that it is seen as a hard planar wall both by the
other solute sphere and by the solvent species.

\subsubsection{Case $\sigma_a=\sigma_b$}
Let us suppose now that the two
solute particles are identical and, for the sake of simplicity, that
the solvent is monodisperse. In that case, Eq.\ \eqref{19} reduces
to
\beq
G_{aa}(s)=\frac{e^{-\sigma_{a} s}}{2\pi
s^2}\frac{L_{aa}(s)\left[1+\alpha
s-A_1(s)\right]+L_{a1}({s})A_{a}(s)}{(1+\alpha s)\left[1+\alpha s -
A_1(s)\right]}.
\label{19.2}
\eeq
The second virial coefficient $B_2$ and the ``stickiness'' parameter
$\tau^{-1}$ associated with the depletion potential $u_{aa}(r)$ are
evaluated in Appendix \ref{appB}. It is shown there that the
depletion potential predicted by the RFA in the colloidal limit
$\sigma_a/\sigma_1\to\infty$ is narrower and \textit{much} deeper
than that predicted by the PY approximation. In fact, the
combination of depth and width represented by the stickiness
parameter $\tau^{-1}$ is divergent in the RFA and finite in the PY.

\subsubsection{Case $\sigma_b/\sigma_a\to\infty$}
 In the limit
$\sigma_b\to \infty$ the solute particle  $b$ is felt as a wall by both a
solvent particle and by the solute particle $a$. Before taking the
limit $\sigma_b\to \infty$, let us introduce the \textit{shifted}
radial distribution function
\beq
\gamma_{ab}(z)=g_{ab}(z+\sigma_{ab}),\quad z\geq 0.
\label{38}
\eeq
In Laplace space,
\beq
G_{ab}(s)=e^{-\sigma_{ab}s}\left[\sigma_{ab}\N_{ab}(s)-\N'_{ab}(s)\right],
\label{39}
\eeq
where
\beq
\N_{ab}(s)=\int_0^\infty dz\, e^{-s z}\gamma_{ab}(z)
\label{40}
\eeq
is the Laplace transform of $\gamma_{ab}(z)$  and $\N'_{ab}(s)=d
\N_{ab}(s)/ds$. In the wall limit $\sigma_b\to\infty$,  Eq.\
(\ref{39}) yields
\beqa
\N_{wa}(s)&=&\lim_{\sigma_b\to\infty}\frac{2}{\sigma_b}
e^{\sigma_{ab}s}G_{ab}(s)\nn &=& \frac{2}{1+\alpha s}\left[
\frac{1}{2\pi s^2}\overline{L}_{aw}(s)+ \sum_{i=1}^N
e^{\sigma_{ai}s}G_{ai}(s)\overline{A}_{iw}(s)\right],\nn
\label{46}
\eeqa
where in the last step we have made use of Eq.\ (\ref{15}) and have
defined
\beq
\overline{L}_{aw}(s)\equiv
\lim_{\sigma_b\to\infty}\frac{L_{ab}(s)}{\sigma_b},\quad
\overline{A}_{iw}(s)\equiv
\lim_{\sigma_b\to\infty}\frac{A_{ib}(s)}{\sigma_b}.
\label{48b}
\eeq
{}From Eqs.\ \eqref{3.7}, \eqref{3.8}, \eqref{3.13}, \eqref{3.14},
and \eqref{3.17} we get
\beq
\overline{L}_{aw}(s)=\overline{L}_{aw}^{(0)}+\overline{L}_{aw}^{(1)}s+\overline{L}_{aw}^{(2)}s^2,
\label{49}
\eeq
\beqa
\overline{A}_{iw}(s)&=&\rho x_i\left[\varphi_2(\sigma_i s)\sigma_i^3
\overline{L}_{iw}^\zero +\varphi_1(\sigma_i s)\sigma_{i}^2
\overline{L}_{iw}^\one \right.\nn &&\left.+\varphi_0(\sigma_{i}
s)\sigma_{i} \overline{L}_{iw}^\two\right],
\label{47}
\eeqa
\beq
\overline{L}_{iw}^\zero=\lambda'-\pi \alpha
\lambda\rho\sum_{j=1}^Nx_j\sigma_j g_{wj}(\sigma_{wj}),
\label{41}
\eeq
\beq
\overline{L}_{iw}^\one=\frac{\lambda}{2}+\frac{\lambda'}{2}\sigma_i-\pi
\alpha \frac{\lambda}{2}\rho\sigma_i \sum_{j=1}^Nx_j\sigma_j
g_{wj}(\sigma_{wj}),
\label{42}
\eeq
\beq
\overline{L}_{iw}^\two=\pi \alpha g_{wi}(\sigma_{wi}),
\label{43}
\eeq
where $i=a,1,\ldots,N$ in Eqs.\ \eqref{41}--\eqref{43}. The
corresponding expressions for the depletion potential and force are
\beqa
\beta u_{wa}(z)&=&-\ln \gamma_{wa}(z)\nn &=&-\ln
{\mathcal{L}^{-1}[\Gamma_{wa}(s)]},
\label{phibis}
\eeqa
\beqa
\beta F_{wa}(z)&=& -
\beta\frac{du_{wa}(z)}{dz}=\frac{\gamma'_{wa}(z)}{\gamma_{wa}(z)}\nn
&=&\frac{\mathcal{L}^{-1}[s
\Gamma_{wa}(s)-\gamma_{wa}(0)]}{\mathcal{L}^{-1}[\Gamma_{wa}(s)]}.
\label{Fdepbis}
\eeqa

In the case of a monocomponent solvent (i.e., $\sigma_i=\sigma_1$,
$i\geq 1$), one has $\overline{A}_{iw}(s)=x_i\overline{A}_w(s)$ and
Eq.\ (\ref{46}) becomes
\beq
\N_{wa}(s)=\frac{1}{\pi s^2}\frac{\overline{L}_{aw}(s)\left[1+\alpha
s-A_1(s)\right]+L_{a1}({s})\overline{A}_{w}(s)}{(1+\alpha
s)\left[1+\alpha s - A_1(s)\right]}.
\label{48}
\eeq
Of course, the same result can be obtained from Eq.\ \eqref{19}.

\section{Results}
\label{sec4} 
\begin{figure}[h]
\includegraphics[width=1.0\columnwidth,angle=0]{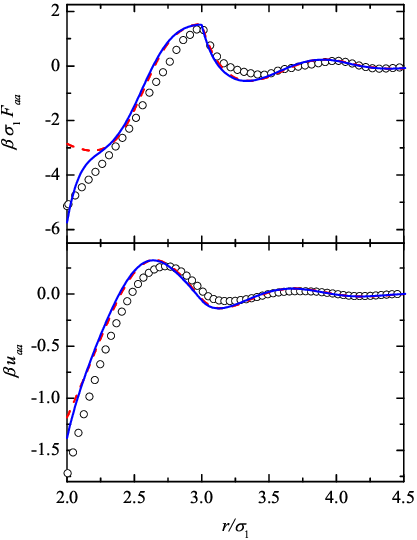}
\caption{(Color online) Depletion force and depletion potential
between two identical (big) hard spheres embedded into a solvent
bath of (small) hard spheres as functions of distance. In this case,
$\Sigma=1$, $R=2$, and $\eta=0.3$. Solid line: RFA approach; dashed
line: PY result;  circles: simulation data from Ref.\
\protect\onlinecite{AL01}.
\label{fig1}}
\end{figure}

\begin{figure}[h]
\includegraphics[width=1.0\columnwidth,angle=0]{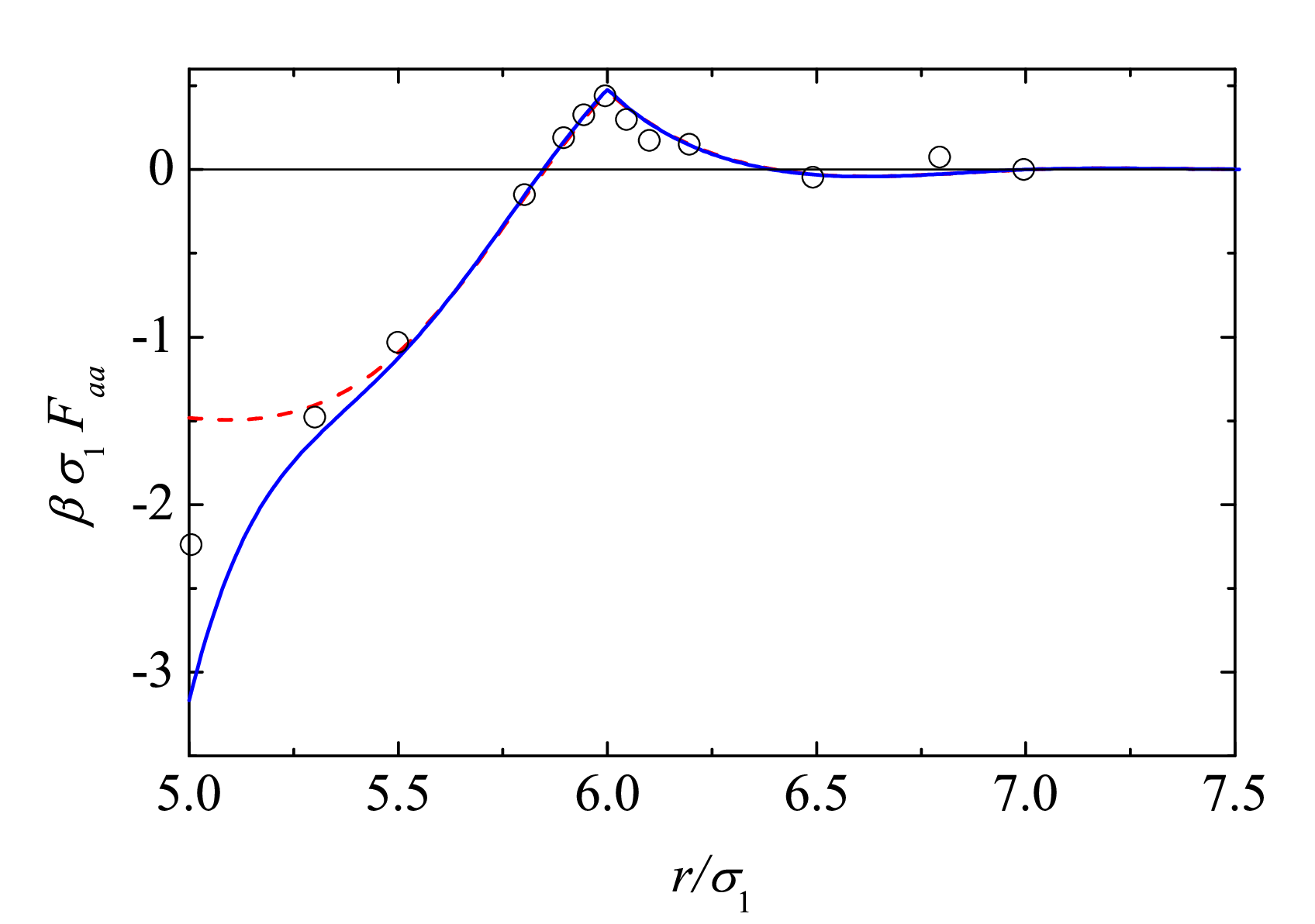}
\caption{(Color online) Depletion force between two identical (big)
hard spheres embedded into a solvent bath of (small) hard spheres as
a function of distance. In this case, $\Sigma=1$, $R=5$, and
$\eta=0.116$. Solid line: RFA approach; dashed line: PY result;
circles: simulation data from Ref.\ \protect\onlinecite{DAS97}.
\label{fig2}}
\end{figure}

\begin{figure}[h]
\includegraphics[width=1.0\columnwidth,angle=0]{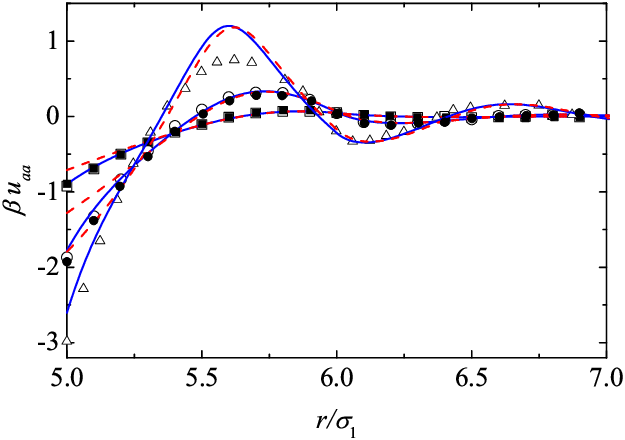}
\caption{(Color online) Depletion potential between two identical
(big) hard spheres embedded into a solvent bath of (small) hard
spheres as a function of distance. In this case, $\Sigma=1$, $R=5$,
and results are displayed for {three} values of $\eta$. Solid lines:
RFA approach; dashed lines: PY results; squares: simulation data for
$\eta=0.1$ from Refs. \protect\onlinecite{MK07} (open symbols) and
\protect\onlinecite{LM04} (filled symbols); circles: simulation data
for $\eta=0.2$ from Refs.\ \protect\onlinecite{MK07} (open symbols)
and \protect\onlinecite{LM04} (filled symbols); {triangles:
simulation data for $\eta=0.3$ from Ref.\
\protect\onlinecite{LM04}}.
\label{fig3}}
\end{figure}

\begin{figure}[h]
\includegraphics[width=1.0\columnwidth,angle=0]{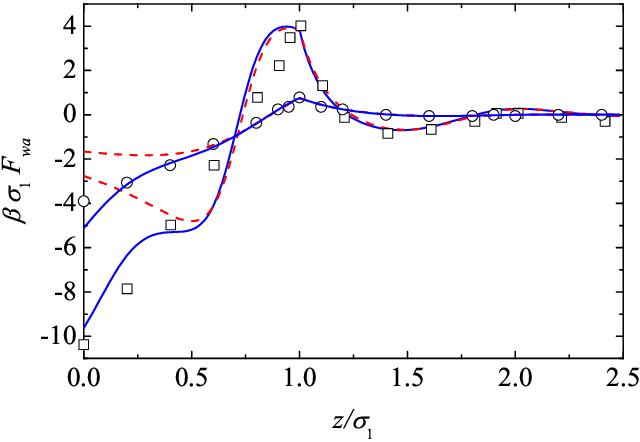}
\caption{(Color online) Depletion force between a hard planar wall
and a (big) sphere in a background fluid of (small) hard spheres. In
this case, $\Sigma \to \infty$, $R=5$, {and results are displayed
for two values of $\eta$}. Solid line: RFA approach; dashed line: PY
result; circles: simulation data {for $\eta=0.1$} from Ref.\
\protect\onlinecite{DAS97}; {squares: simulation data for $\eta=0.2$
from Ref.\ \protect\onlinecite{DAS97}}.
\label{fig4}}
\end{figure}

\begin{figure}[h]
\includegraphics[width=1.0\columnwidth,angle=0]{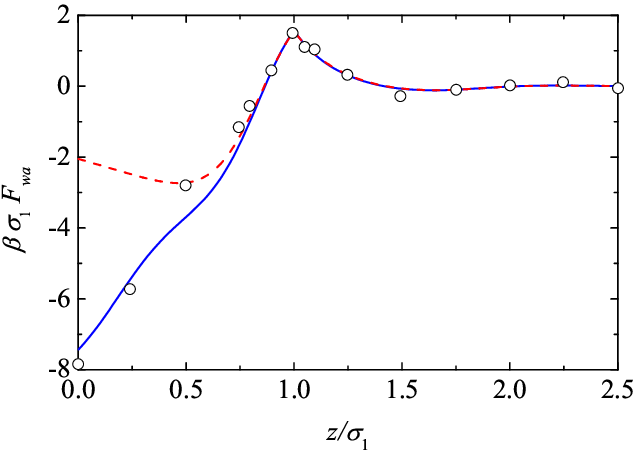}
\caption{(Color online) Depletion force between a hard planar wall
and a (big) sphere in a background fluid of (small) hard spheres. In
this case, $\Sigma \to \infty$, $R=10$, and $\eta=0.1$. Solid line:
RFA approach; dashed line: PY result; circles: simulation data from
Ref.\ \protect\onlinecite{DAS97} \label{fig5}}
\end{figure}

In this section we illustrate the results that one
obtains using our approach    by considering some representative
cases. For simplicity, we will restrict ourselves to a monocomponent
solvent so that $\sigma_i=\sigma_1$ ($i\leq N$). Without loss of
generality we will measure distances in units of $\sigma_1$ and so
the important parameters will be the solvent packing fraction $\eta$
and the size ratios $\Sigma\equiv \sigma_b/\sigma_a$ and $R\equiv
\sigma_a/\sigma_1$. In Figs.\ {\ref{fig1}--\ref{fig5}} we present
the curves obtained using both the PY theory and the RFA approach as
well as the corresponding simulation
data.\cite{DAS97,AL01,LM04,MK07}

As can be seen from the figures, the RFA results certainly represent
an improvement  over the PY theory in all cases {for both the
depletion force and the depletion potential}, yielding in particular
much better values for the {well depth} in the depletion potential.
Our analysis will begin with the cases where both solute particles
have the same size ($\Sigma=1$), namely those in Figs.\
{\ref{fig1}--\ref{fig3}. The RFA results are clearly superior to the
PY ones in the region $\sigma_a\leq r\lesssim
\sigma_a+\frac{1}{2}\sigma_1$. For larger distances, however, the
RFA and PY predictions are hardly distinguishable.} When $R=2$ (see
Fig.\ \ref{fig1}) {the oscillations of both the RFA and the PY
curves are slightly dephased with respect to the simulation data. A
similar behavior is exhibited by the density functional theory shown
in Fig.\ 2a of Ref.\ \onlinecite{AL01}. Figure \ref{fig1} also shows
that the depletion force is a much more stringent quantity than the
depletion potential. In particular, the PY theory predicts a local
minimum of $F_{aa}$ (associated with an inflection point of
$u_{aa}$) at $r\simeq \sigma_a+0.2\sigma_1$, while both the RFA and
the simulation data present a monotonic increase of $F_{aa}$ in the
region $\sigma_a\leq r\leq \sigma_a+\sigma_1$}. If $R=5$ (Figs.\
{\ref{fig2} and \ref{fig3})}, 
we observe a good performance of the RFA for the depletion force at $\eta=0.116$ (Fig.\ \ref{fig2}), except
near contact. On the other hand, the theory is able to capture even quantitatively all the features of the depletion potential for
$\eta=0.1$ and $\eta=0.2$ (Fig.\ \ref{fig3}). For $\eta=0.3$,
paradoxically in contrast to the case $R=2$ of Fig.\ \ref{fig1}, it
follows correctly the trend of the oscillations but otherwise
overestimates the barrier height. Although not shown here and most
likely related with the previous deficiency, also for $\eta=0.3$
the force starts to present features that seem not to occur in the
simulations. We will come back to this point later on in connection
with the hard planar wall limit $\Sigma \to \infty$.

Now we turn to a more stringent situation, namely the case where the
depletion effect takes place between a hard planar wall and a solute
sphere, i.e.,  $\Sigma \to \infty$. In this instance, {as shown in
Figs.\ {\ref{fig4} and \ref{fig5}}, the agreement between the RFA
results and the simulation data is also reasonably good.
Particularly rewarding is the fact that, at least for $\eta=0.1$,
one gets a good performance} even when $R=10$ ({Fig.\ \ref{fig5}}).
{Analogously to the cases with $\Sigma=1$, the RFA strongly improves
over the PY results for distances  $z\lesssim \frac{3}{4}\sigma_1$
from the wall but both theories practically coincide for larger
distances. Also, the PY theory predicts a spurious local minimum of
the depletion force near $z=\frac{1}{2}\sigma_1$}. It should be
pointed out that in the planar wall limit one also starts to get a
peculiar behavior, not shown in the figures, for relatively low
densities ($\eta=0.2$ if $R=10$). {This behavior, also shared by the
PY theory, has similar features to the ones mentioned in connection
with the poorer performance for some systems having $\Sigma =1$,
that is, the appearance of spurious local minima in the depletion
forces and of inflection points in the depletion potentials. While
in the case $\Sigma=1$ within the RFA approach they may have their
origin on the decreasing reliability of the contact values of the
radial distribution functions with increasing size disparity $R$,
when $\Sigma \to \infty$ the features have to do with the fact that
in the hard planar wall limit the radial distribution functions both
for the PY theory and in the RFA may become negative around the
first minimum, \cite{MYSL07} which is clearly unphysical.}

\section{Discussion}
\label{sec5} In this paper we have derived the depletion force and
potential between two (in principle different in size) large spheres
whose interaction is mediated by the presence of a multicomponent
hard-sphere mixture (the solvent) composed of smaller particles.
This has been done by using the RFA approach and the end results
turn out to be completely analytical in Laplace space. {One may say
that the RFA approach not only retains the analytical character of
the PY theory and the good performance of this latter at long
distances, but it has some further assets as well. Hence,  apart
from the elimination of the thermodynamic consistency problem, it is
also able to correct the main drawbacks in the PY formulation in
connection with the present problem, namely the poor prediction of
the short distance behavior and of the well depth in the depletion
potential and the non-divergent character of the stickiness
parameter in the colloidal limit. Since it seems natural that the
most relevant part of the depletion potential be the one
corresponding to short distances, the improvement over the PY result
in this particular region may be considered as a success of the RFA
approach. But the fact that after such an improvement one can also
cater for the (correct) long and intermediate distance behaviors
represents another nice feature of the approach.} It should be clear
that, while for simplicity in the illustrative examples we have
considered that the solvent is a monocomponent fluid, our
development is far more general allowing us in principle to examine
the same problem but with the solvent being a polydisperse
hard-sphere mixture.\cite{HWT03} As far as we know, no simulation
data for such a system are available and so a comparison in this
instance is not possible yet. In general, our expectation that the
RFA produces reasonably accurate results both for the depletion
forces and the depletion potential for low and moderate densities,
provided the solute-solvent size  ratio $R$ is not too big, is
fulfilled.

We have already pointed out one of the limitations of the RFA
approach (also present in the PY theory), namely the fact that in
extreme conditions it may lead to unphysical (negative) values for
the distribution  function $\gamma_{wa}(z)$, which in turn yield
spurious features in the depletion interaction $u_{wa}(z)$. One
technical point must be mentioned at this stage. It concerns the
choice of contact values for the radial distribution functions of
the mixture and the isothermal susceptibility. While here we have
considered the eCSK3 contact values [{see} Eq.\ \eqref{eCSK3}] and
an isothermal compressibility $\chi$ that is thermodynamically
consistent [see Eq.\ \eqref{n2}], the RFA approach does not forbid
the possibility of other choices. For instance, for high values of
$R$ one could instead take the simulation results or the \emph{ad
hoc} proposal of Henderson and Chan\cite{HC97} for the contact value
$\gamma_{wa}(0)$. However, we have checked that, when using the
empirical $\gamma_{wa}(0)$, the region where $\gamma_{wa}(z)$ takes
negative values does not disappear, although those values  become
less   negative. {We would expect of course that the more accurate
the contact values of the radial distribution functions and the
isothermal susceptibility we use as an input, the better the
performance of our development. This, however, remains to be
assessed.}

Finally, we want to point out that in this paper we have restricted
ourselves to the infinite dilution limit of the two solute
particles. This has  allowed us to equate the depletion potential
with the potential of mean force. If the concentration of the solute
is increased, this approximation will cease to be valid. An
important asset of the RFA is that it also yields analytical
expressions for the direct correlation functions and the bridge
functions of the mixture. These expressions could in principle be
used for finite concentrations of the solute, for instance following
the formulation of the depletion potential made by
Casta\~{n}eda-Priego \textit{et al.}\cite{CRM06} This we plan to do
in future work.

\begin{acknowledgments}

We want to thank   J. G. Malherbe, W. Krauth, E. Allahyarov, and H.
L\"{o}wen for kindly providing us with their simulation data. M.
L\'opez de Haro acknowledges the partial financial support of
DGAPA-UNAM under project IN-110406. This work has been supported
 by the Ministerio de
Educaci\'on y Ciencia (Spain) through Grant No.\ FIS2007--60977
(partially financed by FEDER funds) and by the Junta de Extremadura
through Grant No.\ GRU07046.

\end{acknowledgments}

\appendix
\section{Derivation of the exact low density behavior of $g_{ij}(r)$\label{appA}}
In a general mixture, the cavity function corresponding to the pair
$ij$ is defined as $y_{ij}(r)\equiv e^{\beta\phi_{ij}(r)}g_{ij}(r)$,
where $\phi_{ij}(r)$ is the interaction potential. To first order in
density,
\beq
y_{ij}(r)=1+\rho y_{ij}^{(1)}(r)+\mathcal{O}(\rho^2),
\quad y_{ij}^{(1)}(r)=\sum_k x_k y_{ij;k}^{(1)}(r),
\label{A6}
\eeq
where
\beq
y_{ij;k}^{(1)}(r)
= \int d\mathbf{r}' \,f_{ik}(r')f_{jk}(|\mathbf{r}-\mathbf{r}'|),
\label{A1}
\eeq
$f_{ij}(r)\equiv e^{-\beta\phi_{ij}(r)}-1$ being the Mayer function.
In the case of hard spheres, $f_{ij}(r)=-\Theta(r-\sigma_{ij})$, so
that
\beq
y_{ij;k}^{(1)}(r)= V(\sigma_{ik},\sigma_{kj};r),
\label{A7_y}
\eeq
\beq
g_{ij;k}^{(1)}(r)=
V(\sigma_{ik},\sigma_{kj};r)\Theta(r-\sigma_{ij}),
\label{A7}
\eeq
where $V(R_1,R_2;r)$  denotes the intersection volume of two spheres
of radii $R_1$ and $R_2$ whose centers are a distance $r\leq
R_1+R_2$ apart. If $r<R_1-R_2$ (where, without loss of generality,
we have assumed that $R_1\geq R_2$) the small sphere is entirely
contained inside the large one, so that the intersection volume is
just the volume of the small sphere, i.e.,
$V(R_1,R_2;r)=\frac{4\pi}{3}R_2^3$. On the other hand, if
$r>R_1-R_2$, the intersection volume is the sum of the volumes of
two spherical caps of heights $h_1$ and $h_2$, respectively, i.e.,
$V(R_1,R_2;r)=v(R_1;h_1)+v(R_2;h_2)$, where we have denoted by
$v(R;h)$ the volume  of a spherical cap of height $h$ in a sphere of
radius $R$. Its expression is
\beq
v(R;h)=\frac{\pi}{3}h^2(3R-h).
\label{AA1}
\eeq
It remains to obtain $h_1$ and $h_2$ in terms of $R_1$, $R_2$, and
$r$. A simple geometrical construction shows that
\beq
h_1+h_2=R_1+R_2-r,\quad R_1^2-(R_1-h_1)^2=R_2^2-(R_2-h_2)^2,
\label{A3}
\eeq
whose solution is
\begin{widetext}
\beq
h_1=\frac{(R_1+R_2-r)(R_2-R_1+r)}{2r},\quad
h_2=\frac{(R_1+R_2-r)(R_1-R_2+r)}{2r}.
\label{A4}
\eeq
The final result is then
\beq
V(R_1,R_2;r)=\begin{cases} \frac{4\pi}{3}R_2^3,&r<R_1-R_2,\\
\frac{\pi}{12r}(R_1+R_2-r)^2[r^2+2(R_1+R_2)r-3(R_1-R_2)^2],& R_1-R_2<r<R_1+R_2,\\
0,& r>R_1+R_2.
\end{cases}
\label{A5}
\eeq
\end{widetext}

Equation \eqref{11AO} follows from Eqs.\ \eqref{A7} and \eqref{A5},
where it is assumed that $\sigma_{ij}\leq \sigma_{ik}+\sigma_{kj}$
for all sets $\{i,j,k\}$.

\section{Effective stickiness of the depletion
potential\label{appB}} The second virial coefficient associated with
the {depletion} potential $u_{aa}(r)$ is
\beqa
B_2&=&-{2\pi}\int_0^\infty dr\, r^2 \left[e^{-\beta
u_{aa}(r)}-1\right]\nn
&=&\frac{2\pi}{3}\sigma_a^3\left\{1-\frac{3}{\sigma_a^3}\int_{\sigma_a}^\infty
dr\, r^2 \left[g_{aa}(r)-1\right]\right\}.
\label{27}
\eeqa
{}From here one can define the ``stickiness'' parameter\cite{NF00}
\beqa
\tau^{-1}&=&4-\frac{6B_2}{\pi\sigma_a^3}\nn
&=&-\frac{12}{\sigma_a^3} \frac{\partial}{\partial
s}\left[G_{aa}(s)-e^{-\sigma_a s}\frac{1+\sigma_a
s}{s^2}\right]_{s=0}.
\label{28}
\eeqa

Making use of Eq.\ (\ref{19.2}) one  gets an \emph{explicit}
expression of $\tau^{-1}$ in terms of the packing fraction $\eta$,
the size ratio $\sigma_a/\sigma_1$, the RFA parameter $\alpha$, and
the imposed contact values $g_{aa}(\sigma_a)$,
$g_{a1}(\sigma_{a1})$, and $g_{11}(\sigma_1)$. In the special case
of the PY approximation ($\alpha=0$), the result is
\beqa
\tau^{-1}&=&\frac{\eta}{2(1+2\eta)^2}\left[12(1+2\eta)+3(5+4\eta)\sigma_1/\sigma_a\right.\nn
&&\left.+6(1-\eta)(\sigma_1/\sigma_a)^2
+(1-\eta)^2(\sigma_1/\sigma_a)^3\right].
\label{29}
\eeqa
In this approximation, the stickiness parameter $\tau^{-1}$ is lower
bounded by
\beq
\tau_{\text{coll}}^{-1}=\frac{6\eta}{1+2\eta}.
\label{29coll}
\eeq
 In fact, this lower bound is the value in the \textit{colloidal} limit $\sigma_a/\sigma_1\to\infty$.
Moreover, the PY contact values in the colloidal limit are
\beqa
g_{aa}(\sigma_{a})&=&\frac{\sigma_a}{\sigma_1}\frac{3\eta}{2(1-\eta)^2},\nn
 g_{a1}(\sigma_{a1})&=&\frac{1+2\eta}{(1-\eta)^2},\\
g_{11}(\sigma_{1})&=&\frac{1+\eta/2}{(1-\eta)^2}.\nonumber
\label{31PY}
\eeqa

In contrast to Eq.\ \eqref{29coll},  the stickiness parameter
predicted by the RFA ($\alpha\neq 0$) in the colloidal limit becomes
\begin{widetext}
\beq
\tau_{\text{coll}}^{-1}=12(\sigma_1/\sigma_a){\alpha^*}g_{aa}(\sigma_{aa})+6\eta\frac{1+3{\alpha^*}[1-2g_{a1}(\sigma_{a1})+2\eta
g_{11}(\sigma_{1})]-6{\alpha^*}^2g_{a1}(\sigma_{a1})\left[1-(1-\eta)g_{a1}(\sigma_{a1})\right]}{1+2\eta+6{\alpha^*}
\eta\left[1-2(1-\eta)g_{11}(\sigma_{1})\right]},
\label{30}
\eeq
\end{widetext}
where $\alpha^*\equiv \alpha /\sigma_1$. Equation (\ref{30}) implies
that, unless $g_{aa}(\sigma_{a})\sim (\sigma_a/\sigma_1)$, the
stickiness parameter \textit{diverges} in the colloidal limit, i.e.
\beq
\tau_{\text{coll}}^{-1}=12(\sigma_1/\sigma_a){\alpha^*}g_{aa}(\sigma_{a}).
\label{30.2}
\eeq
The behavior $g_{aa}(\sigma_{a})\sim (\sigma_a/\sigma_1)$ appears in
the PY theory.  However, other theories (like the SPT, the BGHLL,
and the one proposed by us in Ref.\ \onlinecite{SYH99}) assume that
$g_{aa}(\sigma_{a})\sim (\sigma_a/\sigma_1)^2$, while
$g_{aa}(\sigma_{a})\sim (\sigma_a/\sigma_1)^3$ in our recent
proposal\cite{SYH05} and $\ln g_{aa}(\sigma_{a})\sim
(\sigma_a/\sigma_1)$ according to Henderson and Chan.\cite{HC97}  A
simple geometrical argument shows that the divergence of $\tau^{-1}$
in the colloidal limit is not an artifact of the RFA. The parameter
$\tau^{-1}$ essentially measures the area (in units of $\sigma_a^3$)
below the curve $r^2[g_{aa}(r)-1]$ between $r=\sigma_a$ and
$r\to\infty$. For large $\sigma_a/\sigma_1$ the range of
$g_{aa}(r)-1$ is expected to be of the order of $\sigma_1$.
Therefore, the area can be estimated as
\beq
\tau_{\text{coll}}^{-1}\sim \frac{1}{\sigma_{a}^3} \sigma_a^2
[g_{aa}(\sigma_a)-1]\sigma_1\sim
\frac{\sigma_1}{\sigma_a}g_{aa}(\sigma_a),
\label{32}
\eeq
in agreement with the leading term in Eq.\ (\ref{30}). To refine
that argument, let us define the range $\xi$ of $r^2[g_{aa}(r)-1]$
as
\beqa
\xi^{-1}&=&-\left.\frac{\partial}{\partial r} \ln\left[r^2
g_{aa}(r)\right]\right|_{r=\sigma_a}\nn
&=&-\frac{g_{aa}'(\sigma_{a})}{g_{aa}(\sigma_{a})}-\frac{2}{\sigma_a}.
\label{33}
\eeqa
Taking into account the definition of $G_{aa}(s)$ as the Laplace
transform of $rg_{aa}(r)$, we have
\beq
\xi^{-1}=-\frac{1}{\sigma_a}-\lim_{s\to\infty} s\left[\frac{s
e^{\sigma_a s}G_{aa}(s)}{\sigma_a g_{aa}(\sigma_{a})}-1\right].
\label{34}
\eeq
In the PY approximation, the result is
\beq
\xi_{\text{coll}}=\sigma_1\frac{1-\eta}{2(1+2\eta)}
\label{34PY}
\eeq
in the colloidal limit. On the other hand, the RFA yields in that
limit
\beqa
\xi_{\text{coll}}&=&\sigma_1\left\{\frac{g_{a1}(\sigma_{a1})}{g_{aa}(\sigma_{a})}\frac{\sigma_a}{\sigma_1}
\frac{3\eta}{1-\eta}\right.\nn
&&\left.+\frac{1}{\alpha^*}\left[1-\frac{\sigma_a}{\sigma_1}\frac{3\eta}{2(1-\eta)g_{aa}(\sigma_{a})}\right]\right\}^{-1}.
\label{35}
\eeqa
If $g_{aa}$ diverges more rapidly than $\sigma_a/\sigma_1$, we get
\beq
\xi_{\text{coll}}=\alpha^*\sigma_1=\alpha,
\label{36}
\eeq
which is generally much shorter than the PY result. In general, we
have
\beq
\tau_{\text{coll}}^{-1}=12\mu\frac{g_{aa}(\sigma_{a})}{\sigma_a}
\xi_{\text{coll}},
\label{37}
\eeq
where $\mu$ is of the order of 1. In the PY approximation we get
$\mu=1/3(1-\eta)$, while $\mu=1$ in the RFA.

\end{document}